\documentclass[twocolumn,showpacs,aps,nofootinbib]{revtex4}
\usepackage{color}   
\usepackage{graphicx}
\usepackage{dcolumn}
\usepackage{bm}

\begin{document}
\def\bbox#1{\hbox{\boldmath${#1}$}}
\def\bb    #1{\hbox{\boldmath${#1}$}}
\def\blambda{{\hbox{\boldmath $\lambda$}}}
\def\eeta{{\hbox{\boldmath $\eta$}}}
\def\bxi{{\hbox{\boldmath $\xi$}}}
\def\bzeta{{\hbox{\boldmath $\zeta$}}}

\title{ Momentum Kick Model Description of the Ridge in $\Delta
  \phi$-$\Delta \eta$ Correlation in $pp$ Collisions at 7
  TeV$^\dagger$
\footnote[0]{${}^\dagger$ Based in part on a talk presented at the
  Workshop on High-pT Probes of High-Density QCD at the LHC at
  Palaiseau, France, May 30 - June 1, 2011.}
}

\author{Cheuk-Yin Wong
 \protect{\thanks{$\dagger$ Based in part on a talk
    presented at the Workshop on High-pT Probes of High-Density QCD at
    the LHC at Palaiseau, France, May 30 - June 1, 2011.}}
}

\affiliation{Physics Division, Oak Ridge National Laboratory, 
Oak Ridge, TN 37831}

\date{\today}

\begin{abstract}

The near-side ridge structure in the $\Delta \phi$-$\Delta \eta$
correlation observed by the CMS Collaboration for $pp$ collisions at 7
TeV at LHC can be explained by the momentum kick model in which the
ridge particles are medium partons that suffer a collision with the
jet and acquire a momentum kick along the jet direction.  Similar to
the early medium parton momentum distribution obtained in previous
analysis for nucleus-nucleus collisions at $\sqrt{s_{{_{NN}}}}$=0.2 TeV,
the early medium parton momentum distribution in $pp$ collisions at 7
TeV exhibits a rapidity plateau as arising from particle production in
a flux tube.
\end{abstract}

                                                                         
\maketitle

\section {Introduction}

Recently at the LHC, the CMS Collaboration observed a $\Delta
\phi$-$\Delta \eta$ correlation in $pp$ collisions at 7 TeV
\cite{CMS10}, and in PbPb collisions at $\sqrt{s_{_{NN}}}=2.76$ TeV
\cite{CMS11}, where $\Delta \phi$ and $\Delta \eta$ are the azimuthal
angle and pseudorapidity differences of two produced hadrons,
respectively.  The correlation appears in the form of a ``ridge'' that
is narrow in $\Delta \phi$ at $\Delta \phi\sim 0$ and $\Delta \phi\sim
\pi$, but relatively flat in $\Delta \eta$.  Similar ridge structures
have been observed previously in high-energy nucleus-nucleus $AA$
collisions at RHIC by the STAR Collaboration
\cite{Ada05,Ada06,Put07,Bie07,Wan07,Bie07a,Abe07,Mol07,Lon07,Nat08,Fen08,Net08,Bar08,Dau08,Lee09,Tra08},
the PHENIX Collaboration \cite{Ada08,Mcc08,Che08,Jia08qm,Tan09}, and
the PHOBOS Collaboration \cite{Wen08}, with or without a high-$p_T$
trigger \cite{Ada06,Tra08,Dau08}.

The CMS observation of the ridge in $pp$ and PbPb collisions raise
many interesting questions.  How do the ridges arise in $pp$ and $AA$
collisions?  Can the ridges in $pp$ and PbPb collisions at LHC and in
$AA$ collisions at RHIC be described by the same physical phenomenon?
If so, what are the similarities and differences? Why is the ridge
yield greatest at 1$<p_T<$3 GeV/c?  What interesting physical
quantities do the ridge data reveal?  How are the ridges associated
with a high-$p_T$ trigger related to the ridges associated with the
minimum-bias pair correlation with no high-$p_T$ selection?

Although many theoretical models have been proposed to discuss the
ridge phenomenon in $AA$ collisions
\cite{Won07,Won08ch,Won08,Won08a,Won09,Won09a,Won09b,Shu07,Vol05,Chi08,Hwa03,Chi05,Hwa07,Pan07,Dum08,Gav08,Gav08a,Arm04,Rom07,Maj07,Dum07,Miz08,Jia08,Jia08a,Ham10}
and $pp$ collisions
\cite{Ham10,Dum10,Wer11,Hwa11,Chi10,Tra10,Tra11a,Arb11,Aza11,Hov11,Bau11,Che10,Dre10,Lev11},
the ridge phenomenon has not yet been fully understood.  Most of the
models deal only with some fragmented and qualitative parts of the
experimental data.  The most successful quantitative comparisons with
experimental data have been carried out in the momentum kick model for
the extensive sets of triggered associated particle data of the STAR
Collaboration, the PHENIX Collaboration, and the PHOBOS Collaboration
--- over large regions of $p_t$, $\Delta \eta$ and $\Delta \phi$ phase
spaces, in many different phase space cuts and $p_T$ combinations,
including dependencies on centralities, dependencies on nucleus sizes,
and dependencies on collision energies \cite{Won07,Won08ch,Won08}.  As
it has been tried and tested successfully for $AA$ collisions at RHIC
in previous analyses from which a wealth of relevant pieces of
information have been obtained, it is of interest to examine whether
the momentum kick model can describe the CMS $pp$ data at 7 TeV --- to
provide answers to the interesting questions we have just posed.

We shall first review the qualitative description of the momentum kick
model in Section II.  We provide additional support for the model in Section III
by showing how the momentum kick model can explain many peculiar and
puzzling features in the minimum-bias correlation data of the STAR
Collaboration \cite{Ada06,Tra08,Dau08}.
We then summarize the quantitative contents of
the model in Section IV.  The determination of the centrality
dependence of the ridge yield necessitates the evaluation of the
number of kicked medium particles along the jet trajectory, which we
describe in Section V.  Section VI provides the numerical analysis of
the ridge yield and the total associated particle distribution in
$pp$ collisions at 7 TeV, for comparison with the CMS data.  In
Section VII, we provide answers to the questions posed in the
Introduction concerning the ridge phenomena in $pp$ and $AA$
collisions. 

\section{Qualitative Description of the Momentum Kick Model}

Soon after the observation of the ridge effect in RHIC collisions, a
momentum kick model was presented to explain the phenomenon
\cite{Won07,Won08ch,Won08,Won08a,Won09,Won09a,Won09b}.  In addition to
providing a semi-quantitative explanation of experimental data over
large regions of $p_t$, $\Delta \eta$, and $\Delta \phi$ phase spaces
in STAR, PHENIX, and PHOBOS experiments, the model serves the useful
purposes of identifying and extracting important physical quantities
that are otherwise difficult to measure.

\begin{figure} [h]
\includegraphics[angle=0,scale=0.50]{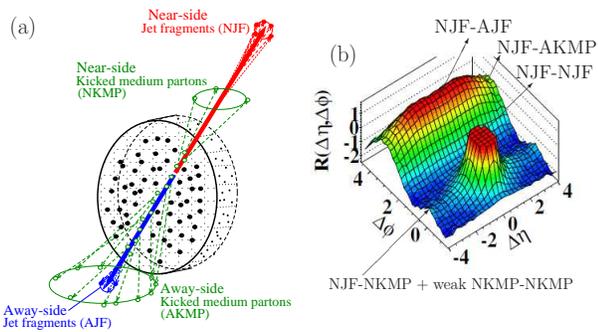}
\vspace*{0.0cm} 
\caption{ (Color online) (a) Schematic representation of the momentum
  kick model.  A jet pair (represented by thick arrows) occur
  back-to-back in a dense medium created in the collision.  The jets
  collide with medium partons, lose energy, and fragment into
  near-side jet-fragments (NJF) and away-side jet-fragments (AJF).
  The near-side kicked medium partons (NMKP) and away-side kicked
  medium partons (AMKP) (represented by open circular points) that are
  kicked by the jets acquire a momentum kick along the jet directions
  and become correlated with the jets at $\Delta \phi\sim 0$ and
  $\Delta \phi\sim \pi$ as associated ``ridge'' particles.  (b) The
  CMS data for high-multiplicity $pp$ collisions at 7 TeV
  \cite{CMS10}. Different regions of the $\Delta \phi$-$\Delta \eta$
  correlations may be identified as correlations of different
  particles associated with the jet fragments (JF) or kicked medium
  partons (KMP).  }
\end{figure}

To understand the physics of the ridge phenomenon, our first task is
to ascertain what the correlated particles are.  We need to specify
the identities of the correlated particles for the case with a high
$p_t$ trigger as well as the minimum-($p_T$)-bias trigger without a
high-$p_T$ selection.  For both cases the two detected particles are
correlated in narrow azimuthal angles at $\Delta \phi \sim 0$ or at
$\Delta \phi \sim \pi$, relative to each other.  Such an experimental
observation suggests that the correlated trigger and associated
particles are related by a collision \cite{Won07}.  It is natural to
consider in the momentum kick model the collisions of jets with medium
partons and attribute the ridge as arising from direct jet-(medium
parton) collisions.  Other models attribute the narrow $\Delta \phi$
correlations to other effects, and future investigations need to sort
out the different consequences in quantitative comparison with
experiment.  Whatever the proposed mechanism may be, collisions of the
jet with the medium partons are bound to occur, and the collisional
correlation between the colliding objects as discussed in the momentum
kick model must be taken into account.

In $pp$ and $AA$ collisions at high-energies, pairwise back-to-back
jets are produced by the hard-scattering process.  These jets
encounter the dense medium that is also created in the collision.  The
jet that encounters less medium material is called the near-side jet.
The other jet that encounters more medium material is the away-side
jet, as depicted in Fig. 1(a). These two jets collide with medium
partons, lose energy, and subsequently break up into jet fragments
(JF) in the form of two narrow back-to-back angular cones.  The
breakup of the near-side jet occurs most likely outside the medium.
The away-side jet is quenched.  The location of the breakup of the
away-side jet, if it is not completely quenched, depends on the degree
of its quenching inside the dense medium.  The medium partons have an
initial momentum distribution at the moment of jet-(medium parton)
collision.  The kicked medium partons (KMP) that are kicked by a jet
acquire a momentum kick along the jet direction.
Subsequently these kicked medium particles materialize as associated
``ridge'' particles to become correlated with the jets in the $\Delta
\phi\sim 0$ and $\Delta\phi \sim \pi$ directions.  The momentum
distribution of the kicked medium particles is given by the initial
momentum distribution displaced by the momentum kick.  As a
consequence, related by jet-(medium parton) collisions are four types
of particles.  They are near-side and away-side jet fragments (JF) and
near- and away-side kicked medium partons (KMP), as shown in
Fig. 1(a).

In our attempt to identify the nature of the correlated particles, it
is important to realize that the observed correlation signals refer to
those above an uncorrelated background, and the $\Delta\phi \sim 0$
and $\Delta\phi \sim \pi$ correlations place severe restrictions which
are satisfied only by causally related particles.  One can pick any
two particles in random. If both particles arise from the bulk medium
that are not related by collisions from the same jet pair as depicted
in Fig. 1, the two-particle correlation will show up as a smooth
background in $\Delta \phi$.  Such a smooth background has been
subtracted from our consideration. The remaining correlation arises
only from the portion of particles that are causally related by
jet-(medium parton) collisions from the {\it same} pair of
back-to-back jets.

Accordingly, we shall identify the two correlated particles as two of
the four types of particles shown in Fig. 1(a).  In the case with a
triggered high-$p_T$ jet particle, the triggered particle of the
$\Delta \phi$-correlated pair can be identified as a near-side jet
fragment (NJF).  The other correlated particle can come from one of
four possibilities:
\begin{enumerate}

\item
 { \bf NJF-NJF correlation}\\ If the other particle is also a
 near-side jet fragment (NJF) from the fragmentation of the same
 near-side jet, they will be correlated in a cone at $( \Delta
 \phi\sim 0, \Delta \eta \sim 0)$.

\item
{ \bf NJF-NKMP correlation}\\ If the other particle is a near-side
kicked medium parton (NKMP), the other particle will be distributed
according to its initial momentum distribution displaced by the
momentum kick.  If the initial momentum distribution of the medium
partons has a rapidity plateau, it will show up as a ridge along $\Delta
\eta$ at $\Delta \phi \sim 0$.

 \item
{ \bf NJF-AKMP correlation}\\ If the other particle is an away-side
kicked medium parton (AKMP), this particle will be distributed
according to its initial momentum distribution displaced by the
momentum kick.  If the initial momentum distribution of the medium
partons has a rapidity plateau, the kicked medium partons will show up
as a ridge along $\Delta \eta$ at $\Delta \phi \sim \pi$.  As there
are more kicked medium partons on the away side than the near side,
the ridge yield will be greater on the away side than the ridge yield
on the near side.  Because of the additional final-state interactions
after the medium partons are kicked, the $\Delta \phi$ and $\Delta
\eta$ distributions are expected to be broadened.  The degree of
$\Delta \phi$ and $\Delta \eta$ broadening increases with the size of
the colliding objects.

\item 
{ \bf NJF-AJF correlation}\\ If the other particle is a member of the
away-side jet fragments (AJF), it will show up as a ridge along
$\Delta \eta$ at $\Delta \phi \sim \pi$.  This $\Delta \eta$ ridge
arises from the longitudinal momnentum difference of the colliding
partons that produce the pair of transverse jets in the
hard-scattering process.  Clearly, because of the multiple collisions
with medium partons as the away-side jet passes through the medium,
the jet becomes more broadly distributed in azimuthal and
pseudorapidity angles.  As a consequence, the away-side jet fragments
AJF that are correlated with the near-side jet have a broader
distribution in $\Delta \phi$ and $\Delta \eta$, and a lower average
transverse momentum. The strength of the NJF-AJF correlation will also
be substantially quenched.  The degree of broadening and quenching
increases with the size of the colliding objects.

\end{enumerate}

The $\Delta \phi$-$\Delta \eta$ correlation observed in $pp$
collisions at 7 TeV by the CMS Collaboration contains these 
distinct features listed above in high-multiplicity events, as
indicated in Fig. 1(b).

It is important to note that the fragmentation of the degraded
near-side jet produces not only high-$p_t$ jet fragments but also
low-$p_t$ jet fragments.  Evidence of the occurrence of low-$p_T$ jet
fragments in the fragmentation of a jet comes from a careful analysis
of the two-particle correlation function of associated particle pairs
down to $p_t$ as low as 2-3 GeV/c \cite{Ada08}.  From the angular
cones of these particles as a function of the $p_t$ of the correlated
pair, one obtains useful systematics on the jet fragments as a
function of $p_t$ \cite{Won09} (See Eqs. (\ref{Njetv}), (\ref{Tjetv}),
and (\ref{ma}) below).  As $p_T$ of the jet fragment decreases, the
jet fragment number $\langle N_{\rm JF}\rangle $ and the jet fragment
temperature $T_{\rm JF}$ decreases, but the jet cone angular width of
the jet fragments increases.  The jet fragment number remains finite
down to very low $p_T$, (even down to $p_T\to 0$) in the systematics.
Another piece of evidence for the occurrence of low-$p_T$ jet
fragments comes from the minimum-bias pair correlation measurements in
STAR \cite{Ada06,Tra08,Dau08} where it was found that there are
clusters of low-$p_T$ particles that are located at $(
\Delta \phi \sim 0, \Delta \eta \sim 0)$ in excess of the background.
These correlations arises from the fragmentation of a parent jet along
the axis of the cone of these correlated pairs.

It should also be realized that the process of fragmentation of a
near-side jet occurs most likely outside the medium as depicted in
Fig.\ 1(a). These low $p_T$ particles are subject to no additional
final-state interactions with the medium.  Furthermore, being kicked
out of the dense interacting medium by the jet, the kicked medium
partons will also be subject to no additional final-state interactions
with the dense medium.  Both the jet fragments and kicked medium
partons can therefore preserve their correlations when they reach the
detectors, resulting in the correlation structures as observed.

Because jet fragments can occur with both high and low $p_T$ values,
we shall generalize the concept of a jet-fragment ``trigger" to
include jet fragments of all $p_T$, both a high-$p_T$ trigger and a
minimum-($p_T$)-bias trigger.  As a minimum-bias trigger contains no
selection in $p_T$, both low-$p_T$ and high-$p_T$ triggers are
possible for a minimum-bias trigger.  Jet fragments from the same jet
correlate with other jet fragments as part of a greater parent jet, no
matter what the $p_t$ values of these two jet fragments may be.  They
are the indicators of the presence of a parent jet.  As indicators of
a jet, they can be used as reference markers to probe the correlation
of other particles that have made collisions with the jet.  By using
such a generalized concept of ``triggered" jet-particles of all $p_T$,
the momentum kick model unifies the description of the observed
``high-$p_T$ triggered" ridge and the ``minimum-$(p_T)$-bias'' ridge
(which is sometimes also called soft ridge).  It is not necessary to
be a high-$p_T$ particle to indicate the presence of a jet, low-$p_T$
particles can also be a jet fragment and indicates the presence of a
jet along the $p_T$ direction, when they are used as correlation
anchors to measure jet effects on other particles.

In spite of these similarities, there is however a notable difference
in (i) the case of a high-$p_T$ trigger and (ii) the case of a
minimum-bias trigger with no high-$p_T$ selection.  In the first case with
a high-$p_T$ trigger, it is reasonable to take this trigger particle
as a near-side jet fragment (NJF), and the near-side correlations come
from its coincidence with another near-side jet fragment (NJF) or with
a kicked medium parton (KMP).  The correlation contains the NJF-NJF,
NJF-NKMP, NJF-AKMP, and NJF-AJF contributions as itemized above.

In the second case of a minimum-bias trigger with a minimum-bias in
$p_T$ selection, one of the two correlated pair particles can be taken
as a ``trigger" and the other as the associated particle.  If this
trigger particle comes from a near-side jet fragment (NJF), the case
of identifying this particle as a jet fragment of all $p_T$ has
already been considered and all earlier considerations apply.  On the
other hand, in the case of a minimum-bias trigger with no bias in
high-$p_T$ selection, low-$p_T$ triggers are possible and the
low-$p_T$ trigger can be a near-side kicked medium parton (NKMP).
Therefore, correlated with this NKMP trigger, there can be additional
NKMP-NKMP, NKMP-AKMP, and NKMP-AJF contributions to the pair
correlations for a minimum-bias trigger which we shall list below.

\begin{enumerate}

\item
{\bf NKMP-NKMP correlation} \\ The associated particle can be a NKMP.
There is thus an additional NKMP-NKMP contribution coming from the
correlation of two near-side medium partons kicked by the same
jet. Each of the kicked medium partons will lie within a small range
of $\phi$ from the jet and therefore the two partons themselves will
be correlated relative to each other, with $\Delta \phi \sim 0$.  If
the initial momentum distribution of the medium partons has a rapidity
plateau, the NKMP-NKMP correlations will show up at $\Delta \phi \sim
0$ as a ridge along $\Delta \eta$ whose range is twice as long as the
ridge in the NJF-NKMP correlation.

\item
{\bf NKMP-AKMP correlation}\\ The associated particle can be an
away-side kicked medium parton (AKMP).  There is an additional
NKMP-AKMP contribution to the two-particle correlation coming from the
correlation of two medium partons kicked by a near-side jet and an
away-side jet of the same hard scattering. The two kicked partons will
be correlated relative to each other with $\Delta \phi \sim \pi$. If
the initial momentum distribution of the two kicked partons has a
rapidity plateau, then the NKMP-AKMP correlation will show up as a
ridge along $\Delta \eta$ at $\Delta \phi \sim \pi$ with a range that
is twice as long as the ridge in the NJF-NKMP correlation.  The
correlation may be attenuated and broadened because of the additional
final-state interactions suffered by the AKMP after it is kicked by
the away-side jet.

\item
{\bf NKMP-AJF correlation} If the low-$p_T$ trigger is a near-side
kicked medium parton and if the away-side jet is not completely
quenched, the associated particle can be an away-side jet fragment
(AJF).  There is the additional NKMP-AJF contribution that shows up as
a ridge along $\Delta \eta$ at $\Delta \phi \sim \pi$.

\end{enumerate}

From the above analysis, we can understand the relationship between
the correlations in the case with a high-$p_T$ trigger and in the case
of a minimum-bias trigger.  If we consider the near side, the case
with a high-$p_T$ trigger involves mainly NJF-NJF and NJF-NKMP
correlations for which the momentum kick model has been successfully
applied to explain the near-side data of the STAR, PHENIX and PHOBOS
Collaborations \cite{Won07,Won08ch,Won08,Won08a,Won09,Won09a,Won09b}.
The case with a minimum-bias trigger involves not only these NJF-NJF
and NJF-NKMP correlations but also additional NKMP-NKMP correlations.
The NKMP-NKMP correlations will contribute only when more than one
medium partons are kicked by the same jet and will be important in the
ridge region.  In extended dense medium as occurs in heavy-ion
collisions, the number of partons kicked by the same jet becomes
considerable and this NKMP-NKMP contribution should be appropriately
taken into account. However, on the near side in $pp$ collisions, the
number of medium partons kicked by the same jet is small, the
NKMP-NKMP contribution is small in comparison with the other NJF-NJF
and NJF-NKMP contributions, and it can be approximately neglected.

\section{Further support for the Momentum Kick Model from STAR Minimum-Bias
Pair Correlation Data}

In addition to the pair correlation measurements with a high-$p_T$
trigger \cite{Ada05,Put07}, the STAR Collaboration has made pair
correlation measurements for the case with a minimum-bias trigger,
with no high-$p_T$ selections \cite{Ada06,Tra08,Dau08}.  The
minimum-bias pair correlation data and the related single-particle
$p_T$ spectrum contain many peculiar and puzzling features that have
so far defied theoretical explanations.  In the momentum kick model,
these features however find simple explanations which provide
additional support for the approximate validity of the momentum kick
model, as indicated below:

\begin{itemize}

\item[(a)]
In the case with a minimum-bias trigger with no high-$p_T$ selections,
as in the STAR measurements in
\cite{Ada06,Dau08,Tra08,Tra10,Tra11a,Tra11b,Tra11c}, the pair
correlations include the NJF-NJF, NJF-NKMP, and NKMP-NKMP correlation
components on the near side, and the NJF-AJF, NJF-AKMP, NKMP-AKMP, and
NKMP-AJF correlation components on the away side, as explained in the
last section.

\begin{figure} [h]
\includegraphics[angle=0,scale=0.5]{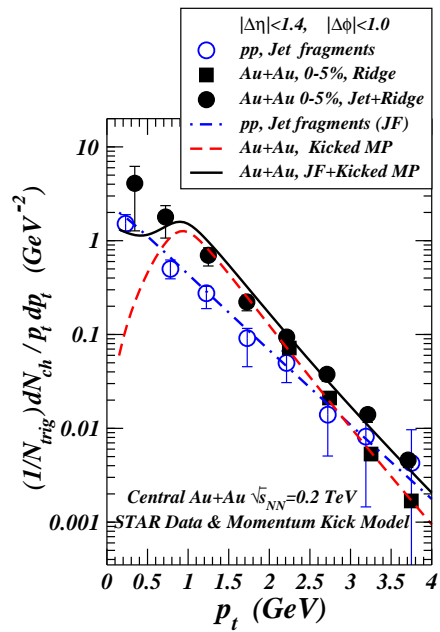}
\vspace*{0.0cm}
\caption{ (Color online)  The $p_T$ distribution of associated
  particles for AuAu collisions at $\sqrt{s_{_{NN}}}$=0.2 TeV as a
  function of $p_T$.  The curves are momentum kick model results and
  the data points are from the STAR Collaboration \cite{Ada05,Put07}  }
\end{figure}

\item[(b)] In the momentum kick model for AuAu collisions at
  $\sqrt{s_{_{NN}}}=$0.2 TeV, the peak of the $p_T$ distribution of
  the medium partons kicked by the near-side jet has been predicted to
  lie at $p_T \sim$ 1 GeV for a high-$p_T$ trigger, as indicated by
  the dashed curve in Fig.\ 2 \cite{Won08ch,Won08,Won08a}.  This
  $p_T$-distribution can be presumed to be also approximately valid
  for medium partons kicked by the away-side jet.  From the relation
  between the $\Delta \phi$ width and the magnitude of the momentum
  kick as shown in Fig. 2 of Ref. \cite{Won07}, we envisage further
  that for the case with a minimum-bias trigger, the undetected
  underlying parent jet (or back-to-back jets) that give rise to the
  narrow $\Delta \phi \sim 0$ or $\Delta \phi \sim \pi$ correlations
  of the kicked medium also provide the same magnitude of the momentum
  kick to the kicked medium partons.  Therefore, the peak of the $p_T$
  distribution of the kicked medium partons will also lie at $p_T\sim$
  1 GeV/c, for the case of minimum-bias trigger with no $p_T$
  selections, on both the near side and the away side.  According to
  the momentum kick model then, the $(p_{T1}, p_{T2})$ correlations of
  the KMP-KMP components on both the near and away sides are expected
  to center at $p_{T1}\sim p_{T2} \sim 1 $ GeV.  They correspond to
  two different kicked medium partons kicked either by the same jet or
  by a jet and its complementary partner on the other side.  They
  acquire the narrow directional correlation of the $\phi$ angles from
  the direction of the jet (or jets) and the $p_T$ momentum kicks from
  the same jet or from the separating back-to-back jets. These KMP-KMP
  correlation components have indeed been observed with
  $(y_{t1},y_{t2})$ correlations centering at $y_{t1}\sim y_{t2}\sim
  2.8$ corresponding to a $(p_{T1}, p_{T2})$ correlations centering at
  $p_{T1}\sim p_{T2}\sim 1 $ GeV \cite{Ada06,Tra08,Dau08}.  The STAR
  Collaboration describes this minimum-bias pair correlation structure
  with various names: the ``hard component", the ``minijet component"
  $H_0$ \cite{Ada06,Tra08,Dau08}, or the ``pQCD component"
  \cite{Tra10,Tra11a,Tra11b,Tra11c}.  From the momentum kick model
  points of view, this pair correlation structure arises naturally as
  the correlation of a pair of medium partons kicked by the same jet
  on the near side or by a jet and its complementary partner on the
  other side.  The momentum kick model includes these pair
  correlations as parts of more general multi-faceted correlations.

\item[(c)]
The STAR data indicate the puzzling feature that the $(p_{T1},
p_{T2})$ pair correlation, centering at $p_T\sim 1$ GeV, persists as a
function of the centrality for both the near-side and the away-side
correlations \cite{Ada06,Tra08,Dau08}.  This indicates that the
detected particles are outside the interaction region of the medium.
For, if they were inside the interaction region, the pair correlation
would have been washed out by additional final-state interactions with
other medium particles as the medium evolves in time, and the
$(p_{T1}, p_{T2})$ pair correlation will not persist as a function of
centrality at the freeze-out point of medium evolution.  This peculiar
feature is consistent with the KMP-KMP correlation components in the
momentum kicked model, where the displacement outside the interaction
region arises naturally and early in the medium evolution as the
kicked medium partons have been kicked out by the jet (or back-to-back
jets) in the jet-(medium parton) collisions.  As the charges of the
medium partons kicked by the same jet or by back-to-back jets are
independent of each other, the $(p_{T1}, p_{T2})$ pair correlation is
the same for the like-sign or unlike-sign pair charges in the momentum
kick model, consistent with the STAR observations \cite{Dau08,Tra10}.

\item[(d)] Being a pair of medium partons kicked by the same jet in
  the NKMP-NKMP correlation component, the magnitude of the pair
  correlation structure on the near-side with $\Delta \phi \sim 0$
  increases as $N_k(b) [N_k(b) -1]$ where $N_k(b)$ is the (average)
  number of medium partons kicked by the jet at the centrality $b$.
  Therefore, there is a threshold of this near-side pair correlation
  component that starts at $N_k(b)\ge 2$.  As the NKMP parton is now
  the trigger particle, the yield per trigger increases as $N_k(b)
  [N_k(b) -1]/N_k(b)$ and thus increases linearly with $ N_k(b) -1$
  after the threshold of $N_k(b)\ge 2$, as observed in the behavior of
  the yield of this pair correlation in the STAR data
  \cite{Ada06,Tra08,Dau08,Tra10}.  The sudden rise of the magnitude of
  the minimum-bias pair correlation (``minijet'') on the near side in
  the STAR data finds a simple explanation.

\item[(e)]
The NKMP-AKMP correlation component involves a pair of medium partons
kicked by a jet and its complimentary partner on the other side with
$\Delta \phi \sim \pi$.  The magnitude of the pair correlation
structure increases as $N_{\rm Nk}(b) N_{\rm Ak}(b) $ where the
subscript ``Nk" is to indicate that $N_{\rm Nk}(b)$ is the (average)
number of medium partons kicked by the near-side jet and the subscript
``Ak" is to indicate that $N_{\rm Ak}(b)$ is the (average) number of
medium partons kicked by the away-side jet.  The momentum kick model
predicts that there will not be as sudden a rise of the magnitude of
the minimum-bias pair correlation for this NKMP-AKMP correlation
component, as compared to the NKMP-NKMP correlation component.

\item[(f)]
The NJF-KMP as well as the KMP-KMP components contain kicked medium
partons that possess their initial momentum distribution before the
kick.  We envisage that the initial momentum distribution of these
medium partons has a rapidity plateau structure and a transverse
momentum temperature parameter $T_{MP}$ that is intermediate between
those of the jet fragments and the bulk medium.  Therefore, the kicked
medium partons retain the rapidity plateau structure but a transverse
momentum distribution displaced by the momentum kick
\cite{Won07,Won08ch,Won08,Won08a,Won09,Won09a,Won09b}.  They show up
as a ridge distribution in the $\Delta \eta$ direction in the $\Delta
\phi$-$\Delta \eta$ pair correlation, for both the near side and the
away side, with the $\Delta \eta$ width much larger than the $\Delta
\phi$ width for these components, as observed in the STAR data
\cite{Ada06,Tra08,Dau08}.  The ridges in the STAR minimum-bias pair
correlation data therefore have a simple explanation.

\item[(g)]
The single-particle distribution will also contain the kicked medium
partons, which show up with a thermal-type distribution with a
temperature $T_{MP}$ displaced by the momentum kick and centering
around $p_T\sim 1$ GeV that is different from the transverse momentum
distribution of the bulk unkicked medium partons.  The single-particle
$p_T$ spectrum therefore contains two components
\cite{Ada06,Tra08,Dau08,Tra10}.  The kicked medium partons constitute
the ``hard component", the ``minijet $H_0$ component''
\cite{Ada06,Tra08,Dau08}, or the ``pQCD component"
\cite{Tra10,Tra11a,Tra11b,Tra11c} of the single-particle spectrum.
Their multiplicities increase with the number of partons kicked by the
jet.  The mean number of partons kicked by a jet depends on the
density of the medium, the jet-(medium parton) cross section, and the
path length of the jet passing through the medium, which in turn
increases with centrality.  Hence, on a per participant pair basis,
the magnitude of this ``hard" component increases with centrality, as
observed in the decomposition of the single-particle spectrum in
\cite{Ada06,Tra08,Dau08,Tra10}.  The STAR data indicates that for the
most central collision, single-particles from this hard component
constitutes a substantial fraction of the total single-particle yield,
implying that jet-(medium parton) collisions are important processes
governing the spatial and momentum distributions of produced particles
in high-energy nuclear collisions.

\end{itemize}

In summary, the momentum kick model gains additional support by
explaining many puzzling features of the pair correlation data in AuAu
collisions at $\sqrt{s_{_{NN}}}=$ 0.2 TeV obtained by the STAR
Collaboration with a minimum-bias trigger.  It is therefore of
interest to examine whether the model can describe the CMS pp data at
7 TeV.

\section{Quantitative Description  of the momentum kick model }

Having presented a qualitative description and supporting evidences
for the momentum kick model, we turn now to the quantitative
description of the model.  To confine the scope of our investigation,
we shall limit our attention to correlations on the near side.  We
shall neglect the NKMP-NKMP contribution, which is a valid
consideration for $pp$ collisions that involve only a small number of
medium partons kicked by the same jet on the near side.

We briefly summarize the main quantitative contents of the momentum
kick model as described in detail in
\cite{Won07,Won08ch,Won08,Won08a,Won09,Won09a,Won09b}.  We follow a jet as
it collides with medium partons in a dense medium and study the yield
of associated particles for a given $p_t^{\rm trig}$ which has been
generalized to include cases of all $p_T$, as providing an angular
location marker for the parent jet.

We label the normalized initial medium parton momentum distribution at
the moment of jet-(medium parton) collisions by $E_i dF/d{\bf p}_i$.
The jet imparts a momentum ${\bf q}$ onto a kicked medium parton,
which changes its momentum from ${\bf p}_i$ to ${\bf
  p}=(p_t,\eta,\phi)={\bf p}_i+{\bf q}$, as a result of the
jet-(medium parton) collision.  By assumption of parton-hadron
duality, the kicked medium partons subsequently materialize as
observed associated ridge hadrons.

The normalized final parton momentum distribution $E dF/d{\bf p}$ at
${\bf p}$ is related to the normalized initial parton momentum
distribution $E_i dF/d{\bf p}_i$ at ${\bf p}_i$ at a shifted momentum,
${\bf p}_i={\bf p}-{\bf q}$, and we have
\cite{Won07,Won08ch,Won08,Won08a,Won09,Won09a,Won09b}
\begin{eqnarray}
\label{final}
\frac{dF}{ p_{t}dp_{t}d\eta  d\phi} 
&=&\left [ \frac{dF}{ p_{ti}dp_{ti} dy_i d\phi_i }
  \frac{E}{E_i} \right ]_{{\bf p}_i ={\bf p}-{\bf q}}
\nonumber\\
& & \times
  \sqrt{1-\frac{m^2}{(m^2+p_t^2) \cosh^2 y}},
\end{eqnarray}
where the factor $E/E_i$ ensures conservation of particle numbers and
the last factor changes the rapidity distribution of the kicked
partons to the pseudorapidity distribution \cite{Won94}.  Changing the
angular variables to $\Delta \eta=\eta-\eta^{\rm trig}$, $\Delta
\phi=\phi-\phi^{\rm trig}$ and characterizing the number of partons
kicked by the jet (per jet) by $\langle N_k \rangle$, we obtain the
charged ridge particle momentum distribution per trigger jet as
\begin{eqnarray}
\label{eq2}
\left [
\frac{dN_{\rm ch}}{ N_{\rm trig} p_t dp_t d\Delta \eta d\Delta \phi } 
\right ]_{\rm ridge}
\!\! \!\!\!\! \!\!\!\!\!\!= \!\!f_R   \frac {2}{3} 
\langle N_k \rangle 
\!\!\!\! \left [ \frac{dF}{ p_{ti}dp_{ti} dy_i d\phi_i } 
     \frac{E}{E_i} \right ]_{{\bf p}_i= {\bf p}-{\bf q}} 
\nonumber\\
 \times
\sqrt{1-\frac{m^2}{(m^2+p_t^2) \cosh^2 y}},~~~~~~
\end{eqnarray}
where $ f_R $ is the average survival factor for produced ridge
particles to reach the detector, and the factor $2/3$ is to indicate
that 2/3 of the produced associated particles are charged.  Present
measurements furnish information only on the product $f_R\langle N_k
\rangle$.  For $pp$ collisions with a small transverse extent, the
kicked medium partons are likely to escape from the interaction region
after the kick and $f_R$ can be approximately taken to be unity.  The
momentum kick ${\bf q}$ will be distributed in the form of a cone
around the trigger jet direction with an average $\langle {\bf
  q}\rangle = q_L {\bf e}^{\rm trig}$ directed along the trigger
direction ${\bf e}^{\rm trig}$ of the parent jet.

We have extracted the normalized initial medium parton momentum
distribution on the right hand side of Eq.\ (\ref{eq2}) from STAR,
PHENIX, and PHOBOS data, we find that the normalized distribution can
be represented in the form
\cite{Won08,Won08ch,Won08a,Won09,Won09a,Won09b}
\begin{eqnarray}
\label{dis2}
\frac{dF}{ p_{ti}dp_{ti}dy_i d\phi_i}&=&
A_{\rm ridge} (1-x)^a 
\frac{ e^ { -\sqrt{m_\pi^2+p_{ti}^2}/T_{\rm MP} }} {\sqrt{m_d^2+p_{ti}^2}},
\end{eqnarray}
where $A_{\rm ridge}$ is a normalization constant,
$x$ is the light-cone variable 
\begin{eqnarray}
\label{xxx}
x=\frac{\sqrt{m_\pi^2+p_{ti}^2}}{m_\pi}e^{|y_i|-y_B},
\end{eqnarray}
$a$ is the fall-off parameter, $y_B$ is the rapidity of the beam
nucleons in the CM system, $T_{MP}$ is the medium parton temperature
parameter, $m_\pi$ is the pion mass, and $m_d=1$ GeV is to correct for
the behavior of the $p_T$ distribution at low $p_T$ as discussed in
\cite{Won08ch,Won08a}.

The total observed yield of associated particles per trigger consists
of the sum of the ridge (NJF-NKMP) component and the jet fragments
(NJF-NJF) component,
\begin{eqnarray}
\label{obs}
\left [ 
\frac{1}{N_{\rm trig}}
\frac{dN_{\rm ch}} 
{p_{t} dp_{t} d\Delta \eta  d\Delta \phi } \right ]_{\rm total}
=
\left [
\frac{dN_{\rm ch}}{ N_{\rm trig} p_t dp_t d\Delta \eta\,d\Delta \phi } 
\right ]_{\rm ridge}
\nonumber\\
+
 f_J \left [ \frac { dN_{\rm jet}^{pp}} {p_t dp_t\, d\Delta \eta\, d\Delta
\phi} \right ]_{\rm JF} \!\!\!\!,  ~~~~
\end{eqnarray}
where $f_J$ is the survival factor of the jet fragments as they
propagate out of the medium.  For fragmentation outside the medium, as
is likely to occur in $pp$ collisions, $f_J$ can be set to unity.  The
experimental associated jet fragment distribution in $pp$ collisions
can be described well by \cite{Won08a}
\begin{eqnarray}
\label{jetfun}
\left [ \frac { dN_{\rm JF}^{pp}} {p_t dp_t\, d\Delta \eta\, d\Delta \phi}\right ]_{\rm JF}
\!\!&=& N_{\rm JF}
\frac{\exp\{(m_\pi-\sqrt{m_\pi^2+p_t^2})/T_{\rm JF}\}} {T_{\rm JF}(m_\pi+T_{\rm JF})}
\nonumber\\
& &\times \frac{1}{2\pi\sigma_\phi^2}
e^{- {[(\Delta \phi)^2+(\Delta \eta)^2]}/{2\sigma_\phi^2} },
\end{eqnarray}
where $N_{\rm JF}$ is the total number of near-side (charged) jet
fragments associated with the $p_T$ trigger and $T_{\rm JF}$ is the
jet fragment temperature parameter.  Extensive sets of data from the
PHENIX Collaboration give $N_{\rm JF}$ and $T_{\rm JF}$ parameters
that vary approximately linearly with $p_t^{\rm trig}$ of the trigger
particle for $pp$ collisions at $\sqrt{s_{_{NN}}}$=0.2 TeV
\cite{Won08},
\begin{eqnarray}
\label{Njetv}
N_{\rm JF} = 0.15 + (0.10  /{\rm GeV/c}) ~p_t^{\rm trig},
\end{eqnarray}
\begin{eqnarray}
\label{Tjetv}
T_{\rm JF}= 0.19 {\rm ~GeV} + 0.06 ~p_t^{\rm trig}.
\end{eqnarray}
We also find that the width parameter $\sigma_\phi$ of the jet fragment cone depends slightly
on $p_t$ which we parametrize as
\begin{eqnarray}
\label{ma}
\sigma_\phi=\sigma_{\phi 0} \frac{m_a}{\sqrt{m_a^2+p_t^2}},
\end{eqnarray}
where $m_a=1.1$ GeV.

With the above formulation, the associated particle distribution is
written in terms of physical quantities, namely, the normalized
initial medium parton momentum distribution $E_i dF/d{\bb p}_i$, the
magnitude of the momentum kick $q_L$, and the number of kicked medium
particles $\langle N_k\rangle$ which depends on the centrality of the
collision.
  
 \section{ Centrality Dependence of the Ridge Yield}

The CMS collaboration obtained the ridge yield as a function of the
charge (particle) multiplicity.  It is necessary to determine the
centrality dependence of both the ridge yield and the charge
multiplicity.

In the momentum kick model, the ridge yield is proportional to the
number of kicked medium partons.  We showed previously how the
(average) number of kicked medium partons per jet, $\langle N_k ({\bb
  b}) \rangle$, can be evaluated as a function of the impact parameter
${\bb b}$ for $AA$ collisions \cite{Won08a,Won09} .  Here, we briefly
summarize these results and apply them to $pp$ collisions by treating
the colliding protons as extended droplets as in the Chou-Yang model
\cite{Cho68}.

\begin{figure} [h]
\hspace*{0.1cm}
\includegraphics[scale=0.45]{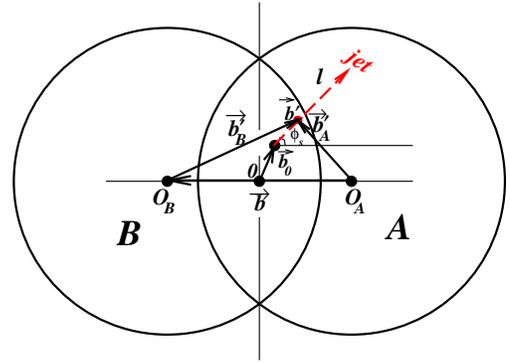}
\caption{ (Color online) The transverse coordinate system used to
  calculate the number of kicked medium partons along the jet
  trajectory in the collision of $A$ and $B$ at an impact parameter
  ${\bb b}={\bb b}_A-{\bb b}_B$.  The jet source point is ${\bb b}_0$
  and the jet-(medium parton) collision point is ${\bb b}'$. The jet
  trajectory lies along ${\bb l}$ and makes an angle $\phi_s$ with
  respect to the reaction plane. }
\end{figure}

Accordingly, we examine the collision of two extended objects $A$ and
$B$ in Fig. 3 and use the transverse coordinate system with the origin
at ${\bb O}$ that is the midpoint between the two centers, ${\bb O}_A$
and ${\bb O}_B$, of the extended objects. In this transverse
coordinate system, the location of the jet production point is labeled
as ${\bb b}_0$, measured from the origin $\bb O$.  The jet is produced
by the collision of a projectile parton and a target parton at ${\bb
  b}_0$.  The jet production point ${\bb b}_0$ measured relative to
the two nucleon centers ${\bb O}_A$ and ${\bb O}_B$ are then given by
\begin{eqnarray}
{\bb b}_A&=&{\bb b }_0+{\bb b}/2,\label{ba} \\
 {\bb b}_B&=&{\bb b}_0-{\bb b}/2. \label{bb}
\end{eqnarray}
From the Glauber model, the probability of finding a target parton at
${\bb b}_0$ is $T_A({\bb b}_A)$, and the probability of finding a
projectile parton at ${\bb b}_0$ is $T_B({\bb b}_B)$.  The probability
for the production of a jet at ${\bb b}_0$ in the collision of $A$ and
$B$ at an impact parameter $\bb b$, $P_{\rm jet}({\bb b}_0,{\bb b})$,
is
\begin{eqnarray}
P_{\rm jet}({\bb b}_0,{\bb b})
&=&
\frac{T_A({\bb b}_0+{\bb b}/2) T_B({\bb b}_0-{\bb b}/2) }{ \int d{\bb b}_0 T_A({\bb b}_0+{\bb b}/2) T_B({\bb b}_0-{\bb b}/2)},
\end{eqnarray}
which is normalized as 
\begin{eqnarray}
\int d{\bb b}_0 P_{\rm jet}({\bb b}_0,{\bb b}) = 1.
\end{eqnarray}

The magnitude of the ridge structure depends on the transverse
momentum of the trigger particle.  The greater is the transverse
momentum of the trigger particle above approximately 10 GeV/c, the
less will be the probability for the appearance of the ridge structure
\cite{Ada08,CMS11}.  There is a dependence of the jet-(medium parton)
interaction on the trigger $p_T$ that needs to be further investigated
in the future.

In our present work, we shall limit our attention to a parent jet that
can interact with medium partons to lead to the ridge structure.  This
means that the parent jet $p_T$ considered is approximately of order
10 GeV with trigger $p_T$ fragments from low $p_T$ values up to a few
GeV/c.  We consider the parent jet to traverse on the near side along
the trajectory $\bb l$ that is measured from the point of production
${\bb b}_0$ and points in the $\phi_s$ direction with respect to the
reaction plane, as shown in Fig. 3.  The jet will collide with medium
partons along its way.  We consider one such a collision at the
transverse coordinate ${\bb b}'$ as measured from the origin ${\bb
  O}$.  The collision point ${\bb b}_A'$ and ${\bb b}_B'$ measured
relative to the two centers ${\bb O}_A$ and ${\bb O}_B$ are then given
by
\begin{eqnarray} 
{\bb b}_A'&=&{\bb b }'+{\bb b}/2,\label{bap} \\
 {\bb b}_B'&=&{\bb b}'-{\bb b}/2, \label{bbp}\\
{\bb b}&=&{\bb b}_A'-
 {\bb b}_B'.
\end{eqnarray}
 The collision point ${\bb b}'$  dependence on 
${\bb b}_0$, $\bb l$,
 and  $\phi_s$   is given by
\begin{eqnarray}
\label{bp}
{\bb b}' ({\bb b}_0, {\bb l},\phi_s)=(b_x', b_y') = (b_{0x}+l\cos
\phi_s,b_{0y}+l\sin \phi_s),
\end{eqnarray}
 which will be needed later on to evaluate the number of kicked medium
 partons.  We label the number of kicked medium partons per jet as
 $N_k({\bb b}_0, \phi_s,{\bb b})$.  The average of $N_k( {\bb
   b}_0,\phi_s, {\bb b})$ with respect to ${\bb b}_0$ is
\begin{eqnarray}
{\bar N}_k(\phi_s,{\bb b} )  & \equiv &  \langle N_k({\bb b}_0,\phi_s, {\bb b}) \rangle_{{\bb b}_0}
\nonumber\\
&=&\frac{\int d {{\bb b}_0} N_k({{\bb b}_0},\phi_s, {\bb b}) P_{\rm jet}({\bb b}_0,{\bb b})}{
 \int d {{\bb b}_0}  P_{\rm jet}({\bb b}_0,{\bb b})}.
\end{eqnarray}

The jet is attenuated along its way, and the attenuation is described
by $\exp\{-\zeta N_k({{\bb b}_0,\phi_s},{\bb b})\}$ with an
attenuation coefficient $\zeta$ that depends on the energy loss and the
change of the fragmentation function on energy \cite{Won08}.  So when
the jet attenuation is taken into account, we get
\begin{eqnarray}
& &{\bar N}_k(\phi_s,{\bb b}) \equiv   \langle N_k({\bb b}_0,\phi_s, {\bb b} ) \rangle_{{\bb b}_0} \nonumber\\
& &  \!\!\!\!=\frac{\int d {{\bb b}_0} N_k({{\bb b}_0},\phi_s, {\bb b}) e^{-\zeta N_k({{\bb b}_0},\phi_s,{\bb b}) } P_{\rm jet}({\bb b}_0,{\bb b}) }
{\int d {{\bb b}_0} e^{-\zeta N_k({{\bb b}_0},\phi_s,{\bb b}) } P_{\rm jet}({\bb b}_0,{\bb b})}.
\end{eqnarray}
We further get the average over angle $\phi_s$ and we get
\begin{eqnarray}
\label{20}
{\bar {\bar N}}_k ({\bb b}) &\equiv & \langle N_k({\bb b}_0,\phi_s,{\bb b} ) \rangle_{{\bb b}_0,\phi_s} \nonumber\\
&=&\frac{1}{\pi/2}\int_0^{\pi/2} d\phi_s 
 \langle N_k({\bb b}_0,\phi_s, {\bb b}) \rangle_{{\bb b}_0}  .
\end{eqnarray}
To proceed further, we need to evaluate $N_k({\bb b}_0,\phi_s,{\bb
  b})$.  The number of jet-(medium parton) collisions along the jet
trajectory that originates from ${\bb b}_0$ and makes an angle
$\phi_s$ with respect to the reaction plane is
\begin{eqnarray}
\label{nkk0}
N_k({\bb b}_0,\phi_s,{\bb b}) = \int_0^{\infty} \sigma \, dl
\frac{dN_{\rm MP}}{dV}  ({\bb b}_A', {\bb b}_B'), 
\end{eqnarray} 
where $0<l<\infty$ parametrizes the jet trajectory, $\sigma$ is the
jet-(medium parton) scattering cross section, and $dN_{\rm MP}({\bb
  b}')/{dV}$ is the medium parton density at ${\bf b}'$ along the
trajectory $\bb l$.  We start the time clock for time measurement at
the moment of maximum overlap of the colliding nuclei, and the jet is
produced by parton-parton collisions at a time $t\sim \hbar/(10 {\rm
  ~GeV})$ which can be taken to be $ \sim 0$. The trajectory path
length $l$ is then a measure of the time coordinate, $t \approx l$.
Due to the longitudinal expansion the density is depleted and the
temperature is decreased as \cite{Bjo83}
\begin{eqnarray}
T \propto (t_0/t)^{c_s^2},
\end{eqnarray}
where $c_s$ is the speed of sound and $t_0$ is the initial time.  As
the entropy density and number densities are proportional to
$T^{1/c_s^2}$, the density of the medium partons therefore varies with
time $t$ as \cite{Bjo83}
\begin{eqnarray}
\label{reg1}
\frac {dN_{\rm MP}}{ dV}(b',t)
=\frac{ dN_{\rm MP}}{ dV} (b',t=t_0) \frac {t_0}{t}.
\end{eqnarray}
The medium parton density $dN_{\rm MP}/dV$ at $(b_{\rm
init}',t=t_0)$ is related to the parton transverse density $dN_{\rm
MP}/d{\bb b}$ at $t_0$ by
\begin{eqnarray} 
\frac {dN_{\rm MP}}{dV} (b',t=t_0)
=\frac {dN_{\rm MP}}{2 t_0d{\bb b}'} (b',t=t_0) .
\end{eqnarray}
To obtain the medium parton density, we introduce the concept of an
extended droplet to describe the proton, as in the Chou-Yang
\cite{Cho68} model with the droplet number of a proton normalized to
unity.  Collisions between droplet elements of one proton and the
droplet elements of the other proton leads to the production of
medium partons.  As in high-energy heavy-ion collision, we assume that
the number of medium partons is proportional to the number of
participating droplet elements $N_{\rm particp}$ with a proportional
constant $\kappa'$
\begin{eqnarray} 
 \frac{dN_{\rm MP}}{dN_{\rm particp}}  =\kappa'. 
\end{eqnarray} 
The initial parton number transverse density ${dN_{\rm MP}}/{d{\bb
    b}'}$ at $t=t_0$ is then related to the corresponding
participating droplet element transverse density ${dN_{\rm MP}}/{d{\bb
    b}'}$ as
\begin{eqnarray} 
\label{eq40}
\frac{dN_{\rm MP}}{d{\bb b}'} =\kappa' \frac{dN_{\rm
particp}}{d{\bb b}'} .
\end{eqnarray} 
We need to evaluate the $\kappa'$  parameter for $pp$ collision at 7 TeV.
Landau hydrodynamical model gives \cite{Lan53,Won08Lc,Won08Ld}
\begin{eqnarray}
N_{\rm ch}= 
K (\xi \sqrt{ s_{_{_{NN}}}}/{\rm GeV})^{1/2}, 
\end{eqnarray}
where $K=2.019$ and $\xi$ is the fraction of incident energy that goes
into particle production in $pp$ and $p\bar p$ collisions.  An
examination of the charge multiplicity in $pp$ and $p\bar p$
collisions indicates that the particle production energy fraction
$\xi$ is approximately 0.5 for these collisions \cite{Bas81}.  So, for
$pp$ at 7 GeV, the average charge multiplicity is
\begin{eqnarray}
N_{\rm ch}=120.
\end{eqnarray}
To determine $\kappa'$, we use a sharp-cutoff thickness function of
the form for the nucleons, (Eq. (12.29) of \cite{Won94})
\begin{eqnarray}
T_A({\bb b}_A) = \frac{3}{2\pi R_A^3} \sqrt{R_A^2 - b_A^2}~~ \Theta(R_A - b_A),
\end{eqnarray}
which gives an average number of participating droplet elements
$\langle N_{\rm particp}\rangle $= 0.4894 and
\begin{eqnarray}
\label{29}
\frac{N_{\rm MP}}{N_{\rm particp}}
=\frac{dN_{\rm MP}}{dN_{\rm particp}}=\frac{3}{2}\frac{dN_{\rm ch}}{dN_{\rm particp}}=  \kappa' =367.
\end{eqnarray}
The transverse participant number density needed in Eqs.\ (\ref{nkk0})
and (\ref{eq40}) along the jet trajectory can be obtained from the
Glauber model to be
\begin{eqnarray}
\label{30}
\frac{dN_{\rm particp}}{d{\bb b}'}({\bb b}_A', {\bb b}_B') =
 [ T_A({\bb b}_A') + T_B({\bb b}_B')  ]~ \Theta ( {\cal R}),   
\end{eqnarray}
where $ \Theta ( {\cal R})$ denotes a step function that is unity
inside the overlapping region and zero outside.  The number of
jet-(medium parton) collisions along the jet trajectory making an
angle $\phi_s$ with respect to the reaction plane is
\begin{eqnarray}
\label{nkk}
\!\!\!\!\!\!
N_k({\bb b}_0,\phi_s,{\bb b}) 
\!\!=\!\! \int_0^{\infty}\!\!\! \frac{\sigma \, dl}{2 t_0}\kappa' 
\left [T_A({\bb b}_A') 
+ T_B({\bb b}_B')\right ] \!  \Theta ( {\cal R})  \frac{t_0}{t},
\end{eqnarray} 
where ${\bb b}_A'$ and ${\bb b}_B'$ are given in terms of ${\bb
  b}_0,\phi_s,{\bb b}$ and $\bb l$ by Eqs.\ (\ref{bap}), (\ref{bbp}),
and (\ref{bp}).  This expression allows us to evaluate the number of
average kicked medium partons per jet ${\bar {\bar N}}({\bb b})$ as a
function of the impact parameter.

There is an amendment which need to be taken into account.  To produce
a medium parton with a transverse mass $m_T\sim p_T$, a period of
initial time $t_0\sim \hbar/p_T$ is however needed to convert the
longitudinal kinetic energy of the collision into entropy so that the
jet-MP collision can commence \cite{Bjo83}.  At RHIC with
$\sqrt{s_{_{NN}}}=$0.2 TeV, the time for producing a particle with a
typical transverse mass or transverse momentum of about 0.35 GeV is
$\hbar/(0.35 {\rm ~GeV/c})\sim 0.6$ fm/c, which is also the time
estimated for the thermalization of the produced matter \cite{Hei02}.
Previous estimates of the jet-MP cross section and attenuation
coefficient $\zeta$ have been obtained with such a $t_0$ value.  At
LHC with $\sqrt{s_{_{NN}}}=$ of 7 TeV, $\langle p_T \rangle$ =0.545 GeV/c
\cite{CMS10a}, which is substantially greater than the average
transverse momentum in RHIC collisions at $\sqrt{s_{_{NN}}}$=0.2 TeV.
Consequently, we need a smaller value of $t_0$ for LHC collisions as
compared to RHIC collisions.

Equation (\ref{nkk}) can be substituted into Eq. (\ref{20}) to allow
the evaluation of the number of ridge particles ${\bar {\bar N}}_k
({\bb b})$ as a function of the impact parameter $b$.

We also need the relation between the charge multiplicity
$N_{\rm ch}({\bb b})$ inside the CMS rapidity window as a function of the
impact parameter ${\bb b}$.  Using Eqs. (\ref{29}) and (\ref{30}), we
obtain
\begin{eqnarray}
& &N_{\rm ch}({\bb b})\nonumber\\
&=&
C_{\rm CMS}
\frac{2}{3} \kappa' \int d{\bb b}'[ T_A({\bb b}'+{\bb b}/2) + T_B({\bb b}'-{\bb b}/2)]  \Theta ( {\cal R}) ,  \nonumber\\
\end{eqnarray}
where $C_{\rm CMS}$ is the fraction of produced particles inside the
CMS rapidity window of $-2.4 < \eta < 2.4$.  Assuming a rapidity
plateau for produced particles, this fraction for the multiplicity of
particles, within the CMS rapidity window in $pp$ collisions 7 TeV
with $y_B=8.91$, is
\begin{eqnarray}
C_{\rm CMS}=\frac{ ({\rm CMS~rapidity~range}) }{2 y_B}=0.269.
\end{eqnarray}
By using these results, we can relate the CMS charge multiplicity
$N_{\rm ch}({\bb b})$ and the average number of kicked medium partons
(per jet) ${\bar {\bar N}}_k ({\bb b})$.

\section{Momentum Kick Model Analysis of $pp$ collisions at 7 GeV}

Previously, the momentum kick model analyses of STAR, PHENIX and
PHOBOS data yield a wealth of useful information.  We learn that the
initial momentum distribution is in the form of a rapidity plateau, as
in the production of particles in a flux tube
\cite{Cas74,Bjo83,Won91,Won94,Won09a}, and the transverse distribution
is in the form of a thermal-type distribution with a medium-parton
temperature $T_{\rm MP}$. The STAR ridge data at $\sqrt{s_{_{NN}}}$=0.2
TeV can be described by a medium parton (MP) momentum distribution of
the form in Eq. (\ref{dis2}) with parameters
\cite{Won08ch,Won08,Won08a,Won09,Won09a}
\begin{equation}
a=0.5,  ~~ T_{\rm MP}=0.5 {\rm ~GeV, ~and~} m_a=1{\rm ~GeV}.
\end{equation}
The magnitude of the momentum kick per collision, $q_L$, was found to
be 0.8-1 GeV/c.  The centrality dependence of the ridge yield can be
described by 
\begin{equation}
\zeta=0.20, ~~ \sigma=1.4{\rm ~ mb},~~ t_0=0.6 {\rm~ fm/c}.
\end{equation}
For the description of the jet fragments, extensive set of PHENIX data
gives systematics of the jet fragments as given in Eqs. (\ref{Njetv}),
({\ref{Tjetv}), and (\ref{ma}).
 
In going from $AA$ collisions at $\sqrt{s_{_{NN}}}$=0.2 TeV to $pp$
collisions at 7 TeV, there are similarities and obvious differences.
The plateau structure of the medium parton distributions in the two
cases are expected to be similar, and the extension of the plateau
should similarly depend on the beam rapidity $y_{_{B}}$ as given in
Eq.\ (\ref{dis2}) and (\ref{xxx}).

We need to know the size of the proton and how the multiplicity
depends on centrality.  The extrapolated $pp$ cross section at 7 TeV
(PDG, 2010) gives \cite{PDG10},
\begin{eqnarray}
\sigma_{\rm tot}(pp)\sim 110 {\rm ~mb},
 ~\sigma_{\rm elastic}\sim 30 {\rm ~mb}.
\end{eqnarray}
Therefore, the $pp$ inelastic cross section at this energy is 
\begin{eqnarray}
 \sigma_{\rm inel}\sim 80 {\rm ~mb}. 
\end{eqnarray}
Sum of $p$+$p$ radii in a $pp$ inelastic collision is then
\begin{eqnarray}
R=\sqrt{80 {\rm ~mb} / \pi}=1.59 {\rm ~fm}=R_A+R_B.
\end{eqnarray}
Therefore, each proton has a radius $R_A=0.8$ fm for inelastic
collisions with the production of particles.

There is however an important difference between RHIC and LHC that
must be taken into account.  For CMS data for $pp$
collisions  at $\sqrt{s_{_{NN}}}$=7 GeV \cite{CMS10a},
\begin{eqnarray}
\label{mb}
\langle p_T \rangle = 0.545 {\rm ~GeV/c}.
\end{eqnarray}
For RHIC collisions at $\sqrt{s_{_{NN}}}=$0.2 TeV,
\begin{eqnarray}
\langle p_T \rangle = 0.39 {\rm ~GeV/c}
\end{eqnarray}
Therefore, the average transverse momentum of produced medium particle
at 7 TeV is enhanced from the average transverse momentum of produced
medium particle at  $\sqrt{s_{_{NN}}}$=0.2 TeV by the factor
\begin{eqnarray}
\frac{\langle p_T \rangle (7 {\rm ~TeV})}
{\langle p_T \rangle (0.2 {\rm ~TeV})}
=\frac{0.545 {\rm ~GeV/c}}{0.39{\rm ~GeV/c}}=1.4.
\end{eqnarray}
Because of this enhancement in the average $p_T$ values, it is
necessary to scale those quantities that are related directly to
transverse momentum by this empirical factor of 1.4.  Accordingly, the
relevant parameters that we need to change are the medium parton
temperature $T_{\rm MP}$, the jet fragment temperature $T_{\rm JF}$, and 
the medium parton initial time $t_0$ which varies roughly as
$1/p_T$.  For the analysis of CMS data at 7 TeV, we are well advised
to scale up $T_{\rm MP}$ and $T_{\rm JF}$  by a factor of 1.4 to result in 
\begin{equation}
T_{\rm MP}=0.7 {\rm~ GeV,}
~~T_{\rm JF}=0.266 {\rm~GeV~}+0.084 p_t^{\rm trig}, 
\end{equation}
and reduce the initial time $t_0$ by a factor of 1.4 to get
\begin{equation}
t_0=0.43 {\rm~fm/c}.
\end{equation}

\begin{figure} [h]
\includegraphics[angle=0,scale=0.41]{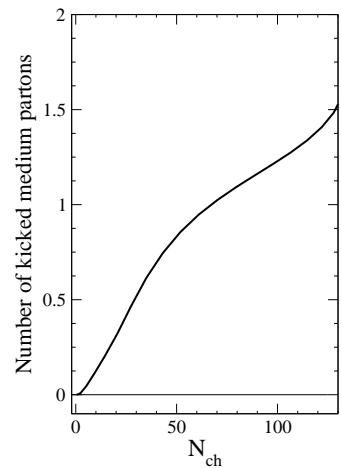}
\vspace*{0.0cm}
\caption{ The number of average kicked medium partons $N_k({\bb b})$
  per minimum-bias trigger particle as a function of charge
  multiplicity $N_{\rm ch}({\bb b})$.  }
\end{figure}
As a function of the impact parameter, we calculate the average number
of kicked medium partons per jet, which gives the ridge yield per jet
as in Eq.\ (\ref{eq2}). After calibrating the number of average
droplet participants with the average number of produced charged
particles as given in Eq.\ (\ref{29}), we calculate the charge
multiplicity as a function of the impact parameter.  These two
calculations gives the ridge yield per jet as a function of the charge
multiplicity.

The jet trajectory calculation indicates that the average number of
kicked partons for the most central $pp$ collision is about 1.5 as
shown in Fig.\ 4.  With the parameters properly scaled according to
the transverse momenta, the only free parameter is $q_L$, the
magnitude of the momentum kick acquired by the medium parton per
jet-MP collision.  Previously, for the STAR data at
$\sqrt{s_{_{NN}}}$=0.2 TeV, the magnitude of $q_L$ was found to be
0.8-1.0 GeV/c per kick.  We vary $q_L$ to fit the variation of the CMS
ridge yield data in different $p_T$ windows.  If the magnitude of
$q_L$ remains at 1 GeV/c, the ridge yield will be too large in the
regions of $0.1<p_T<1$ GeV/c and too small in the region $2 < p_T < 3$
GeV/c. The results do not agree with data.  It is found that the
magnitude of the kick $q_L=2$ GeV/c gives results that are
qualitatively consistent with the data.  The yield per minimum-bias
trigger particle for different regions of associated particle $p_T$,
as a function of the multiplicity, is shown in Fig.\ 5 where the CMS
data are shown as solid points and the momentum kick model results are
shown as curves.  The general trend of the experimental data is
reasonably reproduced by the momentum kick model.

\begin{figure} [h]
\includegraphics[angle=0,scale=0.42]{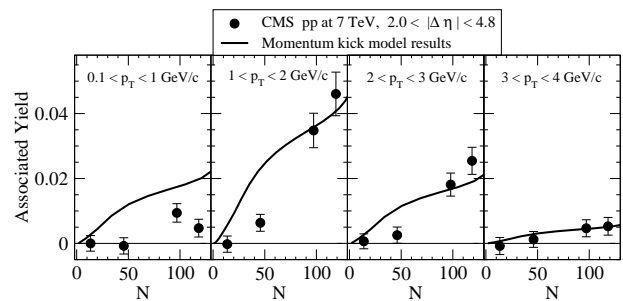}
\vspace*{0.0cm}
\caption{ The ridge yield per minimum-bias trigger particle for
  different regions of associated particle $p_T$, as a function of the
  multiplicity in the interval $2<| \Delta \eta|<4.8$.  (a) is for
  $0.1<p_T<1$ GeV/c, (b) for $1<p_T<2$ GeV/c, (c) for $0.2<p_T<3$
  GeV/c, and (d) for $3<p_T<4$ GeV/c. }
\end{figure}

Another indication of the dependence of the angular distribution of
the associated particle yield on transverse momentum of the associated
particle is shown in Fig. 6 where Fig.\ 6(a) is for $0.1 < p_T < 1.0$
GeV/c, and 5(b) is for $1<p_T<3$ GeV/c.  These total associated
particle yields for the minimum-bias pair correlation has been
calculated for a triggered particle with $p_t^{\rm trig}=0.545$ GeV/c,
corresponding to the minimum-bias average $p_T$ of detected hadrons
given in Eq.\ (\ref{mb}).  The top of the distributions have been
truncated to show the distributions in a finer scale.  The ridge
structure is almost imperceptible for $0.1 < p_T < 1.0$ GeV/c in
Fig.\ 6(a) but shows up clearly for $1.0 < p_T < 3.0$ GeV/c in
Fig.\ 6(b).  The theoretical associated particle yield pattern of the
$\Delta \phi$-$\Delta \eta$ angular distributions as a function of
$p_T$ agrees with those observed in CMS experiments.

\begin{figure} [h]
\includegraphics[angle=0,scale=0.8]{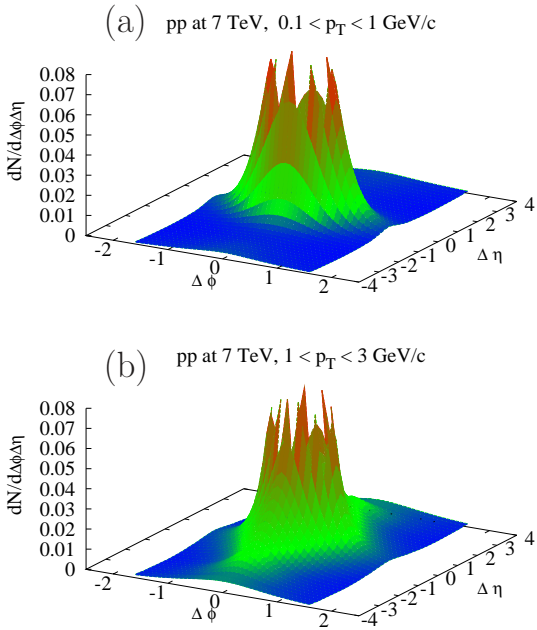}
\vspace*{0.0cm}
\caption{ (Color online) The distribution of associated particles for
  $pp$ collisions at 7 TeV calculated in the momentum kick model.  The
  peaks of the distributions have been truncated at the top.  Fig. 5(a) is for $0.1 < p_T < 1 $ GeV/c and Fig. 5(b) is for $1 < p_T < 3
  $ GeV/c.}
\end{figure}

To exhibit further the dependence of the ridge yield on $p_t$ and
collision energy, we note in Fig.\ 2 that for AuAu collisions at
$\sqrt{s_{{_{NN}}}}$=0.2 TeV within the acceptance windows of the STAR
Collaboration, the ridge yield $dN_{\rm ch}/N_{\rm trig} p_T dp_T$ has
a peak at about 1 GeV/c, which is about the same as the magnitude of
the momentum kick $q_L=1$ GeV/c.  Because $T_{\rm MP} < T_{\rm JF}$,
the NJF-NJF component dominates over the NJF-NKMP components at large
$p_T$.

To make meaningful comparison with $pp$ collisions at 7 TeV, we
calculate $dN_{\rm ch}/N_{\rm trig} p_T dp_T$ in the momentum kick
model with $q_L=2$ GeV/c within the CMS experimental windows with a
$p_T=5$ GeV/c trigger.  The theoretical results are shown as curves in
Fig. 7.  The ridge yield distributions are the medium parton
distributions displaced by the momentum kick along the jet direction.
As a consequence, the peak of the ridge yield (the dashed curve) moves
to a larger value of $p_T$ and becomes broader over a large region
between 0.5 to 2.5 GeV/c for $pp$ collisions at 7 TeV with $q_L=2$
GeV/c.  The large pseudorapidity window for the trigger jet enhances
the broadening of the peak of the distribution for $pp$ collisions.
These theoretical results explain why the associated particle yield is
distributed more in the region of 1$<p_T<3$ GeV/c than in other
regions in $pp$ collisions at 7 TeV.  Note again that because the the
higher temperature for the NJF-NJF component, the ridge yield become
smaller than the NJF-NJF yields as $p_T$ increases above 3 GeV/c.

\begin{figure} [h]
\includegraphics[angle=0,scale=0.4]{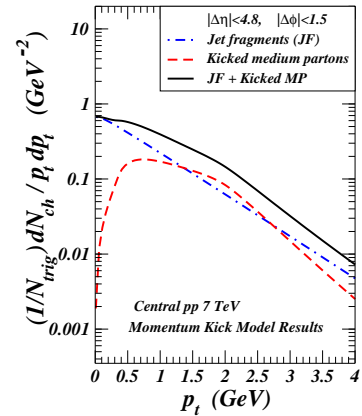}
\vspace*{0.0cm}
\caption{ (Color online) The $p_T$ distribution of associated
  particles for central pp collisions at 7 TeV obtained in the
  momentum kick model.  }
\end{figure}

\section{Conclusions and Discussions}

The CMS observation of the ridge structures in $pp$ collisions at 7 TeV
raises many interesting questions which we can try to answer in the
momentum kick model.  We find that in $AA$ as well as in $pp$
collisions, the ridge arises from the medium partons that are kicked
by the jet and they acquire a momentum kick along the jet direction.
Because the collision occurs at the early stage in the presence of the
jet, the ridge provides direct information on the momentum
distribution of the medium partons at the moment of jet-(medium
parton) collision.

The STAR, PHENIX, and PHOBOS data for $AA$ collisions at RHIC and the
CMS data for $pp$ collisions at LHC are consistent with the initial
momentum distributions in the form of a rapidity plateau, with a
thermal like transverse momentum distribution.  While the functional
form of the momentum distribution is the same, the longitudinal and
transverse momentum at the two energies need to be properly scaled
according the their differences in collision energies.  The
longitudinal rapidity need to be scaled according to the collider beam
rapidity $y_B$ in the center-of-mass system, and the transverse
momentum distribution temperature parameters for $pp$ collision at 7
TeV need to be scaled up by a factor of 1.4 compared to the case of
$\sqrt{s_{_{NN}}}$=0.2 TeV, and the medium particle initial time $t_0$
reduced by the same factor of 1.4.  This factor of 1.4 was estimated
from the ratio of the average transverse momenta of $pp$ collisions at
the two different energies.

We estimate that the average number of kicked medium partons in the
most central $pp$ collisions at 7 TeV with the highest multiplicity is
approximately 1.5.  This is less than the number of kicked medium
partons of about 4 for the most central Au-Au collisions at
$\sqrt{s_{_{NN}}}$=0.2 TeV.  The medium produced in $pp$ collision at 7
TeV is not as dense as the medium produced in the most central AuAu
collisions at $\sqrt{s_{_{NN}}}$=0.2 TeV, but the medium is nonetheless
dense enough for the jet to kick the medium partons to turn them into
ridge particles for our examination.

Using our knowledge of the physical quantities in the momentum kick
model analyses for $AA$ collisions at RHIC energies, there is only a
single parameter we vary in trying to understand $pp$ data at 7 TeV.
We find that in $pp$ collisions at 7 TeV, experimental data suggest a
greater momentum kick value, $q_L=2$ GeV/c, which is twice as large as
$q_L\sim1$ GeV/c for RHIC $AA$ collisions at  $\sqrt{s_{_{NN}}}$=0.2 TeV.

The ridge yields for $pp$ collisions are more prominent in the region
of $1<p_T<3$ GeV/c.  Such a feature arises because the ridge particles
are just particles whose momenta are shifted by the momentum kick.
The shift of $q_L=$2 GeV/c will place the center of the momentum distribution
in the $p_T$ region between 1 and 3 GeV/c region.  Hence the ridge yield
is greater in this region compared to other $p_T$ regions.

The approximate validity of the momentum kick model raises the
interesting question on the nature of the scattering between the jet
and the medium partons.  It suggests a picture of the medium parton
absorbing a part of the jet longitudinal momentum in its scattering
with the jet.  We can envisage that the jet at this stage is a
transversely broad object in the form of a cloud of gluons propagating
together in a bundle, and the complete absorption of a part of the
gluon cloud by the medium parton imparts the longitudinal momentum to
the medium parton to carry the medium parton out to become a ridge
particle.  In this simple description, the jet behaves like a
composite object and the scattering between the jet and the medium is
not a simple two-body elastic scattering process, as in the case with
a jet parton of ultra-high $p_T$.  It is more like ``shooting a water
hose on a bunch of fast-moving ping-pong balls'' \footnote[1]{ The author
thanks Prof. R. L. Ray for such a colorful description of the
momentum kick model.}  Further theoretical and experimental
investigations on the nature of the jet-(medium parton) collision for
relatively low-$p_T$ jets at this stage will be of great interest.

Transverse hydrodynamical expansion will lead to azimuthal
correlations \cite{Shu07,Vol05,Gav08,Gav08a,Dum10,Ham10,Wer11}.
However, a quantitative analysis of the effects of the dynamical
expansion needs to be examined carefully as the transverse flow
depends sensitively on time \cite{Bay83} and it is much slower than
the longitudinal expansion \cite{Lan53}.  Furthermore, the transverse
hydrodynamics for a fluid system with non-isotropic momentum
distribution in the early history of the expansion has not been worked
out in details.  In the analogous case with an isotropically
thermalized fluid, the picture of transverse expansion \cite{Bay83}
reveals that soon after the stage of transverse overlap at which the
jets are produced, the medium is essentially at rest with little
transverse expansion. The transverse expansion commences only after
the rarefaction wave passes through the medium from the outer surface
inward.  The time for the rarefaction waves to travel depends on the
radius of the medium and the speed of the rarefaction wave which is
the speed of sound.  Future investigations on hydrodynamical solutions
for a non-isotropic momentum distributions and azimuthally asymmetric
shapes will provide a more accurate calculation of the effects of
transverse flow in azimuthal correlations.

Whatever the proposed mechanism may be, collisions of the jet with the
medium partons are bound to occur, and the collisional correlation
between the colliding objects as discussed in the momentum kick model
must be taken into account because these collisional correlations will
contribute to the two-particle correlation function.

The momentum kick model provides a unifying description for ridges
with or without a high-$p_T$ trigger.  The description generalizes the
trigger to include high-$p_T$ and low-$p_T$ particles owing to the
fact that jet fragments are found in high $p_T$ as well as in low
$p_T$.  Many peculiar and puzzling features of the minimum-bias
correlation data of the STAR Collaboration
\cite{Ada06,Tra08,Dau08,Tra10} find simple explanations in the
momentum kick model.  If we limit our attention to the near side, the
case with a high-$p_T$ trigger involves mainly NJF-NJF and NJF-NKMP
correlations while the case of minimum-bias pair correlation with
low-$p_T$ triggers involves not only these NJF-NJF and NJF-NKMP
correlations but also the NKMP-NKMP correlation.  The additional
NKMP-NKMP correlation in the low-$p_T$ trigger case will contribute
only when more than one medium partons are kicked by the same jet and
will be important in the ridge region.  In extended dense medium as
occurs in heavy-ion collisions, the number of partons kicked by the
same jet becomes considerable and this NKMP-NKMP contribution should
be appropriately taken into account.

In conclusion, the ridges in both $AA$ collisions at
$\sqrt{s_{_{NN}}}$=0.2 TeV and $pp$ collisions at 7 TeV in LHC can be
described by the same mechanism of the momentum kick model involving
the collision of jets with medium partons.  The momentum distributions
of the medium partons at the moment of jet-(medium parton) collisions
have similar features of a rapidity plateau and a thermal type
transverse momentum distribution.

For $pp$ collisions at 7 TeV, it will be of interest to carry out
measurements of the $(p_{T1},p_{T2})$ correlations with a minimum-bias
trigger that will reveal whether there is a correlation at $p_{T1}
\sim p_{T2} \sim $ 2 GeV, as suggested by the results of Fig.\ 7 and
similar to the correlation of $p_{T1} \sim p_{T2} \sim $ 1 GeV/c for
$pp$ collisions at 0.2 TeV.  Future experiments also call for the
measurement of the $p_T$-distribution of various components with a
high-$p_T$ trigger to separate out different correlation
contributions.  Further acquisitions of more correlation data of the
ridge yields, in large regions of the phase space in different phase
space cuts and combinations, will provide useful information on the
correlation mechanism and the momentum distribution of medium partons
at the early history of the collision.

\centerline{\bf Acknowledgment}
\vspace*{0.3cm} The author wishes to thank Drs. Vince Cianciolo and
R. L. Ray for helpful discussions.  This research was supported in
part by the Division of Nuclear Physics, U.S. Department of Energy.

\end{document}